# Chemical Vapor Deposition Growth of Boron-Carbon-Nitrogen layers from Methylamine Borane Thermolysis Products

Fabrice Leardini[1,2]*, Eduardo Flores[1], Andrés R. Galvis E.[1], Isabel Jiménez Ferrer[1,2], José Ramón Ares[1], Carlos Sánchez[1,2], Pablo Molina[1,2], Herko P. van der Meulen[1,2], Cristina Gómez Navarro[3], Guillermo López Polin[3,4], Fernando J. Urbanos[5], Daniel Granados[5], F. Javier García-García[6], Umit B. Demirci[7], Pascal G. Yot[8], Filippo Mastrangelo[9], Maria Grazia Betti[9], Carlo Mariani[9]

[1] Departamento de Física de Materiales, Universidad Autónoma de Madrid, Campus de Cantoblanco E-28049, Madrid, Spain
[2] Instituto Nicolas Cabrera, Universidad Autónoma de Madrid, Campus de Cantoblanco E-28049, Madrid, Spain
[3] Departamento de Física de la Materia Condensada, Universidad Autónoma de Madrid, Campus de Cantoblanco E-28049, Madrid, Spain
[4] Present address: School of Physics and Astronomy, University of Manchester, Oxford Road, M13 9PL, Manchester, United Kingdom
[5] Instituo Madrileño de Estudios Avanzados en Nanociencia (IMDEA-Nanociencia), Cantoblanco, E-28049, Madrid, Spain
[6] ICTS-Centro Nacional de Microscopía Electrónica, Universidad Complutense de Madrid, E-28040, Madrid, Spain
[7] IEM (Institut Europeen des Membranes), UMR5635 (CNRS, ENSCM, UM), Universite de Montpellier, Place Eugene Bataillon, CC047, F-34095, Montpellier, France
[8] Institut Charles Gerhardt Montpellier (ICGM) UMR 5253 CNRS-UM-ENSCM, Universite de Montpellier, CC 15005, Place Eugene Bataillon, F-34095, Montpellier, France
[9] Dipartimento di Fisica, Università di Roma 'La Sapienza', I-00185 Roma, Italy

* Corresponding author: **fabrice.leardini@uam.es**

**Abstract**

This work investigates the growth of B-C-N layers by chemical vapor deposition using methylamine borane (MeAB) as single-source precursor. MeAB has been synthesized and characterized, paying particular attention to the analysis of its thermolysis products, which are the gaseous precursors for B-C-N growth. Samples have been grown on Cu foils and transferred onto different substrates for their morphological, structural, chemical, electronic and optical characterizations. The results of these characterizations indicate a segregation of h-BN and Graphene-like (Gr) domains. However, there is an important presence of B and N interactions with C at the Gr borders, and of C interacting at the h-BN-edges, respectively, in the obtained nano-layers. In particular, there is significant presence of C-N bonds, at Gr/h-BN borders and in the form of N doping of Gr domains. The overall B:C:N contents in the layers is close to 1:3:1.5. A careful analysis of the optical bandgap determination of the obtained B-C-N layers is presented, discussed and compared with previous seminal works with samples of similar composition.

**Introduction**

In the last years there has been a huge interest in the growth and characterization of few- or single-layered compounds. Among them $MoS_2$, $MoSe_2$, $TiS_3$, Graphene (Gr) and hexagonal boron nitride (h-BN) layers are some of the most investigated systems. In particular Gr and h-BN present many similarities, due to their isoelectronic and isostructural nature, showing a small lattice mismatch (<2%). However, owing to the different polar character of B-N with respect to C-C bond, they also present remarkable differences. Gr is a semimetal with zero bandgap, whereas h-BN is an insulator with a bandgap of about 6 eV. Ternary borocarbonitride (B-C-N) layers have been predicted to behave as semiconductors with adjustable bandgap by tuning B, C and N contents [1]. Compared to transition metal chalcogenides, B-C-N are based on relatively abundant and cheap elements. Moreover, B-C-N present excellent thermal and chemical resistances, showing poor oxidation tendency even at elevated temperatures. These features make few layered B-C-N very promising for 2D device applications. In addition, these compounds present interesting properties for electrocatalysis, such as the Oxygen Reduction Reaction (ORR) [2], the Hydrogen Evolution Reaction (HER) [3] or photocatalytic water splitting [4,5].

The growth of homogeneous B-C-N layers is however hurdled by the strong tendency of the ternary system towards segregation of Gr and h-BN domains [1,6]. This is mainly driven by the fact that C-C and B-N bonds are much stronger than C-N and C-B ones. Many efforts have been devoted to the growth of homogeneous B-C-N layers by using different synthetic approaches, such as chemical vapor deposition (CVD) of different precursors containing B, C and N atoms. In the seminal works [7-9], several precursors containing C and B-N separately (such as $CH_4$ and $NH_3BH_3$) were employed. In this way the obtained layers presented segregation of Gr and h-BN domains.

Recently several groups have investigated the use of single-source precursors containing B, C and N atoms in the same molecules, aiming at the achievement of homogeneous B-C-N layers. Some examples are dimethylamine borane ($C_2H_6NHBH_3$, DMAB hereafter) [10], trimethylamine borane ($C_3H_9NBH_3$, TMAB hereafter) [11] and bis-BN cyclohexane ($B_2N_2C_2H_{12}$) [12], which have been used as single-source precursors for B-C-N growth by CVD. Here, we report on the growth of B-C-N by using a novel molecular precursor which is closely related to them, namely methylamine borane ($CH_3NH_2BH_3$, MeAB hereafter). MeAB, DMAB and TMAB are chemical derivatives of ammonia borane containing one, two and three methyl groups bonded to N, respectively. Besides, the thermolysis of MeAB releases cyclic molecules containing B, C and N atoms, which are in this sense somewhat similar to bis-BN cyclohexane.

Until today the obtaining of ternary homogeneous B-C-N with semiconducting properties remains elusive. In particular, not all previous works give enough evidences of the chemical state and chemical bonding of C, N and B atoms. In this work the obtained samples have been thoroughly characterized by means of a plethora of techniques, aiming at the determination of their morphological, structural, chemical, electronic and optical properties. On the other hand, particular attention has been devoted to the analysis of UV-vis optical absorption measurements to ascertain the optical bandgaps of h-BN and C-rich domains. We show how the claiming of a bandgap opening in C-rich domains induced by B and N doping reported in some previous works was based on an erroneous analysis of the UV-vis optical spectra. A description of the correct models to obtain the optical bandgaps is presented, discussed and compared with previous works.

**Experimental Methods**

*Synthesis and characterization of Methylamine Borane precursor*

MeAB was synthesized in a Schlenk line by reacting sodium borohydride (Sigma Aldrich, 98%, ref. 452882-100G) and methylamine hydrochloride (Sigma Aldrich, ref. M0505-100G) in anhydrous tetrahydrofuran (THF, Acros Organics, 99.5%, Extra Dry), following the route reported in the literature [13-15]:

$$NaBH_4 + (H_2NCH_3)\cdot HCl \rightarrow NaCl + H_2 + BH_3NH_2CH_3 \qquad (1)$$

In a typical experiment, to a suspension of sodium borohydride (0.1 mol) in anhydrous THF (200 mL), methylamine hydrochloride (0.1 mol) was added and the reaction mixture was stirred (550 rpm), at 30 ºC, for 24 h. Subsequently, the residuals solids were filtered off and the solvent removed *in vacuo*, yielding MeAB as a white crystalline solid. MeAB samples were stored and handled in a glove box (MBraun, $H_2O$ < 5 ppm).

The obtained samples were characterized by means of X-ray Powder Diffraction (XRPD). XRPD measurements were performed at room temperature (RT) using a PANalytical X'PERT multipurpose diffractometer using a monochromatic Cu-K$_{\alpha 1}$ source ($\lambda$ = 1.54056 Å) with operating voltage of 40 kV and a beam current of 40 mA. The powders were introduced into a glass capillary tube of 0.5 mm diameter into a glove box (Jacomex P-Box, $H_2O$ < 5 ppm). The capillaries were sealed before experiments to keep the loaded samples out of air and water. For all the diffraction data collected, the unit cell parameters were determined using DICVOL06 [16] and the structure were refined from a structural model obtained using FoX [17] and Jana 2006 software package [18]. Rietveld refinement was carried out using a manually defined background and a Pseudo-Voigt as profile function with a cut-off fixed equal to 12×FWHM. A total of 25 parameters were refined with the B-N and C-N distances restrained to 1.68 Å and 1.54 Å respectively. The asymmetry was corrected using divergence and the atomic displacement parameters were isotropically refined.

The purity and molecular structure of the sample was verified by $^1$H Nuclear Magnetic Resonance (NMR) spectroscopy on a Bruker AVANCE-300 (probe head dual $^1$H/$^{13}$C, 300.13 MHz, 30 °C). Acetonitrile-d3 CD3CN (Eurisotop) was used as solvent and 1 mg of sample was dissolved in this solvent for analysis. The molecular structure of the sample was also analyzed by solid-state $^{11}$B Magic Angle Spinning (MAS) NMR on a Varian VNMR400 ($^{11}$B 128.37 MHz, −10 °C, 18500 rpm).

*Thermolysis of MeAB*

The thermolytic decomposition of the MeAB precursor was investigated by means of Differential Scanning Calorimetry (DSC), Thermogravimetric Analysis (TGA) and Thermal Programmed Desorption-Mass Spectrometry (TPD-MS). Typically a few milligrams of sample were hermetically closed in Al pans inside the glove box and transferred to the DSC and TGA devices. A small hole was done in the caps of the Al pans just before loading the samples into the DSC or TGA instruments. In this way, the exposure time to air is reduced to a few seconds and the possible influence of sample aging is minimized. DSC measurements were done in a calorimeter Q100 from TA Instruments under an Ar flow of 50 sccm. TGA measurements were recorded in a TGA Q500 system from TA Instruments, under an Ar flow of 50 sccm. Concomitant MS analyses of the gases released during TGA runs have been done in a quadrupole mass spectrometer (Pfeiffer, mod. ThermoStar), by collecting the gases evolved from the sample through a silica capillary heated at 190 ºC. Thermolysis experiments have been performed both under a constant heating rate of 2 ºC/min and under isothermal conditions at 90 ºC.

*Growth of B-C-N layers by CVD*

The layers were grown in a quartz tube of 1 m in length and 20 mm in diameter placed inside a cylindrical furnace and connected to a Ar+$H_2$ gas line. The Ar and $H_2$ flows were fixed by using MKS Mass-Flo$^R$ controllers. Cu foil substrates (Alfa Aesar, ref. 46986) with typical dimensions of 10x15x0.025 mm$^3$ were firstly washed with acetone and then with ethanol in an ultrasonic bath for 5 min and then placed in the central zone of the furnace. In order to remove surface oxides in the Cu foils these were heated at 1050 ºC under Ar+$H_2$ flow (50 sccm Ar + 50 sccm $H_2$) during 1 h and then cooled down to 1000 ºC for CVD growth (under 50 sccm Ar). Typically 10 mg of MeAB precursor were loaded in a small glass boat having a steel bar (completely sealed inside the glass) which allows its displacement along the quartz tube with the help of an external magnet. Initially, the MeAB powders were placed downstream at a distance of 30 cm of the furnace (at nearly RT). Once the Cu foils were ready for CVD growth, the MeAB powders were approached to the furnace and heated at 80-90 ºC using the temperature gradient produced along the quartz tube. After 60 min of CVD growth the precursor was cooled down and the samples were quenched to RT by rapidly displacing the tube furnace (which was mounted in a rail line).

*Transfer of B-C-N layers to different substrates*

The obtained B-C-N layers grown on Cu foils were transferred to different substrates in a clean room at IMDEA Nanociencia facilities. The layers grew on both sides of the Cu foils, so we choose the layer on the side facing up and we remove the backside by reactive ion etching by using the following parameters: flowing 15 sccm $O_2$ at a total pressure of 100 mTorr and applying 100 W of power during 60 s. Then, the layers on the Cu substrate were covered with polymethyl methacrylate (PMMA) by spin coating at 5000 rpm during 1 min and baked at 180 ºC for 2 min to dry the PMMA layer. Afterwards the Cu foils were chemically etched overnight by floating the samples (with the PMMA side facing up) in a dilute solution of ammonium persulfate (APS, 0.2 g/ml). Once the Cu was etched the samples were rinsed three times in deionized (DI) water (Milli Q grade) and then deposited onto the desired substrate. In order to eliminate the water present between the layer and the substrate, the samples were purified by keeping them under dynamic vacuum at $10^{-2}$ mbar for several hours. Finally, to remove PMMA residues we tried several methods (heating the samples at 300-500 ºC in $N_2$ or $H_2$ flow, immersing in acetone, etc.). Best results were obtained by immersing the samples a few minutes in dichloromethane at 40 ºC. The transfer of the B-C-N layers to the TEM grids were done with and without using PMMA. Using PMMA facilitates the transfer process, but PMMA residues were difficult to remove completely at the nanometer scale and therefore hinder the analysis by Transmission Electron Microscopy (TEM) and Electron Energy Loss Spectroscopy (EELS). Therefore, we also tried the transfer without using PMMA, which complicates the transfer process (in particular the rinsing with DI water to remove APS residues, since the ultrathin layers where difficult to be seen floating in water and scooped out). It has been observed that rinsing with DI water after the chemical etching process is fundamental, otherwise sulfate residues are observed in TEM analysis. After several trials, some samples were successfully transferred to the TEM grids without using PMMA.

*AFM measurements*

The B-C-N layers were transferred onto a Si substrate. AFM images were acquired in non-contact mode with a homemade microscope controlled with a Dulcinea Control Unit (Nanotec) and WSxM software [19]. Silicon AFM probes from Nanosensors with a nominal force constant of 40 N/m, resonant frequency of ~ 350 kHz and tip radius of 20 nm were used for the measurements.

*TEM and EELS characterization*

The obtained B-C-N layers were transferred to Cu and Cu + C (amorphous) grids for TEM and EELS characterization. A JEOL GRAND ARM transmission electron microscope operated at 80 kV was used. Recording of the EELS spectra were done by using a dedicated ENFINIUM SE spectrometer. Selected Area Diffraction Patterns (SAED) have been also recorded with the TEM apparatus.

*X ray Photoelectron Spectroscopy measurements*

The X-ray Photoelectron Spectroscopy (XPS) measurements were carried out at the LoTUS surface physics laboratory (Sapienza, University of Rome, Italy) in an Ultra High Vacuum (UHV) chamber, with a base pressure in the low $10^{-10}$ mbar range. Photoelectrons were excited by an $Al_{K\alpha}$ photon source (hv = 1486.7 eV), they were measured with a hemispherical electron analyser (VG Microtech Clam-2) used in constant pass energy (PE) mode set at 50 eV, with an energy resolution of 1 eV, further details are available in [20,21]. The electron binding energy (BE) was calibrated by acquiring after each measurement the Au $4f_{7/2}$ core-level set at 84.0 eV BE. The CVD-grown sample on Cu and transferred onto high conductivity Si crystal was afterward air-transferred and mounted in the XPS ultra-high-vacuum system. The sample was annealed up to 500 °C in UHV in sequential steps, and after each step XPS spectra were acquired at RT.

*Raman characterization*

Raman spectra have been acquired in a WITec ALPHA 300AR instrument using a confocal microscope with different lenses (20x and 100x). A laser with excitation wavelength of 532 nm and a power of 1 mW has been used. Additional Raman spectra have been also recorded in different instruments at excitation wavelengths of 488 nm and 325 nm. Raman spectra at 488 nm excitation wavelength were carried out in a diffraction grating spectrometer (0.5 m) with an electron multiplied Peltier Cooled Si-CCD. Spectra were acquired in a confocal set up with a X40 plan-apo microscope objective with a N.A. of 0.65. Excitation power was carried below 2.5 mW. Raman spectra at 325 nm excitation wavelength and 1mW excitation power, were carried out in a diffraction grating spectrometer (1 m) with a liquid nitrogen cooled CCD (Horiba), the emission was collected with a 50x microscope objective with a NA of 0.55.

*Optical Absorption Spectroscopy*

The absorption spectra have been obtained with a double beam UV/vis/NIR Lambda 1050 PerkinElmer spectrometer. The samples were measured in transmission configuration using a spot size of 5 mm$^2$ in the 190-860 nm spectral range. Baselines for the instrument calibration have been recorded before the measurements.

**Results and discussion**

**Characterization of Methylamine Borane**

The obtained MeAB compound has been characterized by X-Ray Powder Diffraction (XRPD) as shown in Figure 1. Diffraction peaks have been indexed using two phases: a main phase of MeAB (space group *Pnma*) and traces of NaBH$_4$ (space group *F-43m*). An additional peak with low intensity is observed at 18.0º. This peak cannot be indexed neither to MeAB nor to sodium borohydride or methylamine hydrochloride reactants, and it could be due to traces of another impurity in the obtained powder. Figure 1 also shows the result of the refinements carried out using the Rietveld method on the MeAB and NaBH$_4$ phases. Lattice parameters of MeAB obtained by Rietveld refinement of the diffraction data are a = 11.1637(3) Å, b = 6.74602(12) Å, c = 4.97017(11) Å. The following agreement factors have been obtained after refinement: GoF = 1.72, Rp = 5.54; Rwp = 7.46. All the details used for the refinement procedure are given in the Supplementary Data (Table S1). As seen on Figure 1 the calculated pattern (red line) fits quite well the experimental data (black line). The unit cell parameters and atomic positions are in very good agreement with those reported in the literature [13]. As concerns the secondary phase of $NaBH_4$, the obtained lattice parameter is a = 6.1588(3) Å, which is also in agreement with previous values [22]. The relative amounts of the MeAB and $NaBH_4$ phases obtained by the refinement of XRPD data are 95.46(7) wt% and 4.54(11) wt%, respectively.

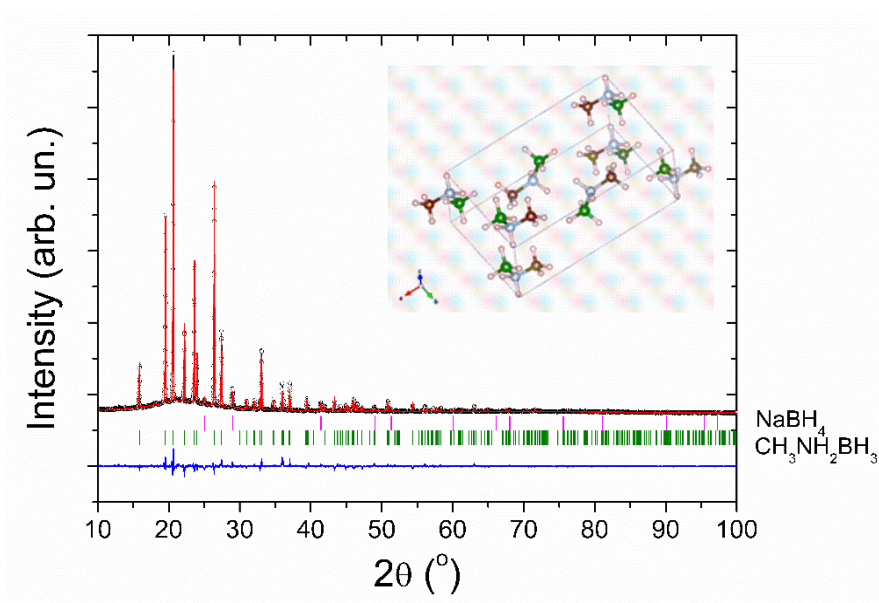

**Figure 1.** Observed (in black) and calculated (in red) powder X-ray diffraction profile for the Rietveld refinement of the obtained MeAB samples. The bottom curve (in blue) is the difference plot on the same scale intensity and the tick marks are the calculated angles for the Bragg peaks in 2θ for MeAB (green) and $NaBH_4$ (pink) (λ = 1.54056 Å). The inset shows the unit cell of MeAB phase.

Further characterizations of MeAB based on Nuclear Magnetic Resonance measurements (both $^1H$ NMR in the liquid state and $^{11}B$ MAS NMR in the solid state) as well as on FTIR spectroscopy have been done, and the results are included in the Supplementary Data (Figs.S1 and S2 and Table S2). These characterizations confirm the successful synthesis of MeAB.

**Thermolytic decomposition of Methylamine Borane**

MeAB is a crystalline solid at ambient conditions. In order to use it as a precursor for the growth of B-C-N layers it must be heated to produce gaseous species that can be used in CVD growth. The reaction scheme of MeAB thermolysis reported in previous works seems to be sensitive to the experimental conditions (flowing gas or static conditions, pressure, temperature, etc.) [13,23,24]. With this in mind the thermal decomposition of MeAB precursor has been investigated by means of DSC and TGA under the same conditions than those used in CVD growth (flowing Ar at ambient pressure). The composition of the gaseous species released during MeAB thermolysis has been analyzed by concomitant MS measurements.

Figure 2 shows the DSC-TGA-MS profile of MeAB recorded at a constant heating rate of 2 ºC/min. The DSC signal (Figure 2a) shows an endothermic peak at 55.3 ºC. This event occurs without significant mass loss and gas release, therefore it can be ascribed to the melting of MeAB (the heat of fusion obtained from the present data is $\Delta H_F$ = 11.9 kJ/mol). Upon subsequent heating a second endothermic event is observed in the temperature range between 100 and 135 ºC. This process is accompanied by a substantial mass loss, as observed in Figure 2b. Concomitant MS characterizations of the gas released during TGA runs have been done in the m/z range 1 to 140 a.m.u. The mass spectrum of the gas released at the temperature at which release rate is maximum is shown in the Supplementary Data (Fig.S3). This spectrum can be properly described by considering the superposition of the MS spectra of $H_2$, MeAB and a mixture of trimethylborazine isomers (cyclic borazine rings bonded to three methyl groups, having a chemical formula $(CH_3NBH)_3$). Accordingly, our results indicate that several competing processes are taking place during MeAB thermolysis: $H_2$ release (Figure 2c), MeAB sublimation and the release of cyclic trimethylborazine molecules (Figure 2d). .

It is seen that the release of $H_2$ starts before that of B-C-N-H molecules, thus confirming that it is linked to a different process, namely to the polymerization of MeAB molecules, which is in competition with their sublimation and the release of trimethylborazine isomers. On the other hand, it must be emphasized that the B-C-N-H gaseous species released during MeAB thermolysis are the precursors used in the subsequent CVD growth of the B-C-N layers, following the procedure described in the Experimental Section.

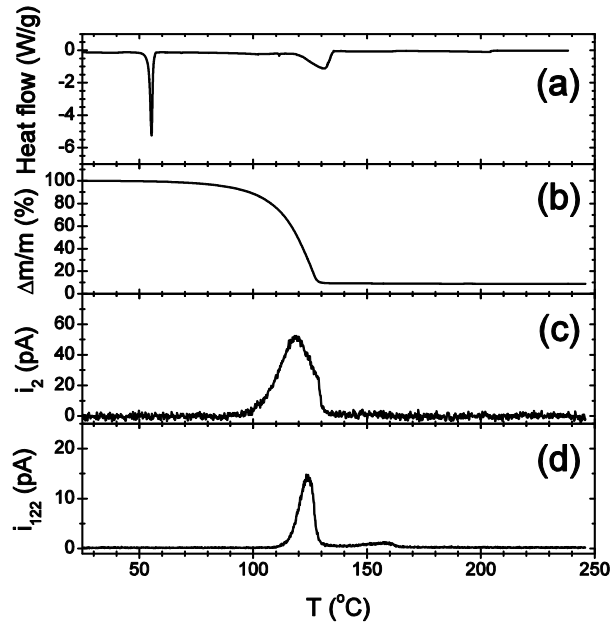

**Figure 2.** Thermal decomposition profiles of MeAB investigated by concomitant DSC-TGA-MS measurements recorded at a constant heating rate of 2 °C/min under Ar flow. (a) DSC signal; (b) TGA profile; (c) MS signal at m/z = 2 corresponding to $H_2$ evolved from the sample. (d) MS signal at m/z = 122 corresponding to trimethylborazine isomers evolved from the sample.

**Characterization of B-C-N layers grown by CVD**

Figure 3 shows the morphology of a typical B-C-N layer transferred onto a Si substrate as observed by AFM. This image shows a good homogeneity at the micrometer scale.

The analysis of the topography distribution indicates that the average sample thickness is of 11±2 nm. The growth of multilayered samples indicates the strong tendency of B, C and N atoms to react and deposit at the B-C-N surface, which acts as a catalyst for subsequent CVD growth.

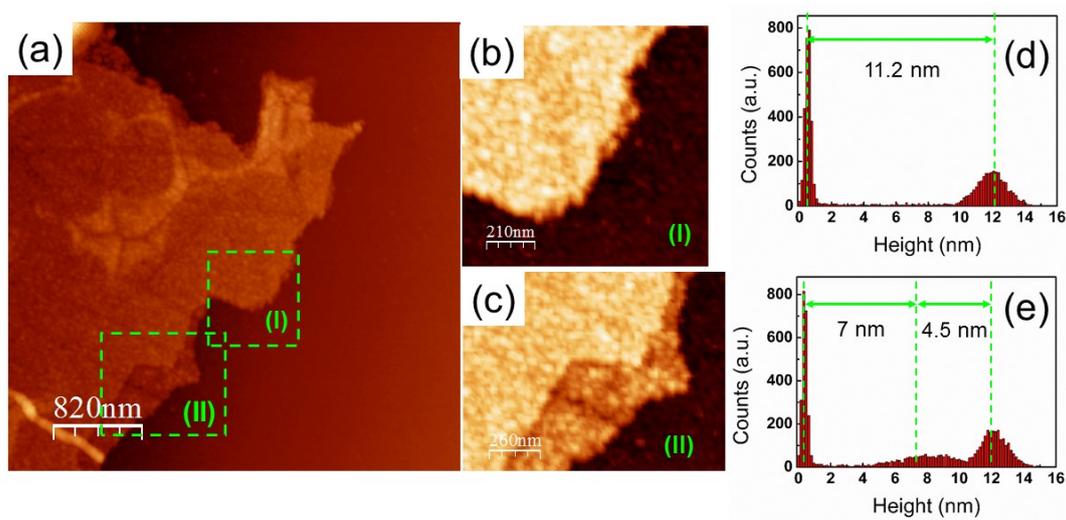

**Figure 3.** AFM images of a B-C-N layer transferred onto Si substrate. The left panel (a) shows a general view of the sample near the edge of the film, with the Si substrate at the right part of the image. The regions marked as (i) and (ii) are shown with higher resolution in the middle panels (b and c, respectively). The topography distributions of these images are shown in the right panels (d and e, respectively). It must be noticed that the zones with heights ranging between 0-1 nm correspond to the Si substrate, whereas the zones with higher heights correspond to the B-C-N layer.

To further investigate the morphology and the chemical bonding properties of the obtained layers, these have been characterized by TEM and EELS. A typical TEM image of the layer can be seen in Figure 4. The sample presents a homogeneous flat morphology. SAED patterns (Figure 4) present two diffraction rings with some reminiscence of a hexagonal symmetry. Interplanar distance obtained from the internal diffraction ring is equal to 2.1 Å, which is ascribed to first order reflections of {100} planes in the hexagonal structure [25]. The external diffraction ring corresponds to an interplanar spacing of 1.2 Å, due to diffraction on {110} planes. The observation of continuous diffraction rings in SAED patterns indicates the polycrystalline nature of the B-C-N layers. The real space explanation is done in terms of the lack of order between adjacent crystalline nano-domains [26] and by disordered stacking of multiple hexagonal layers [27]. Sample thickness calculated from the zero energy loss EELS peak is of ~10 nm, a value which is in good accordance with the AFM measurements.

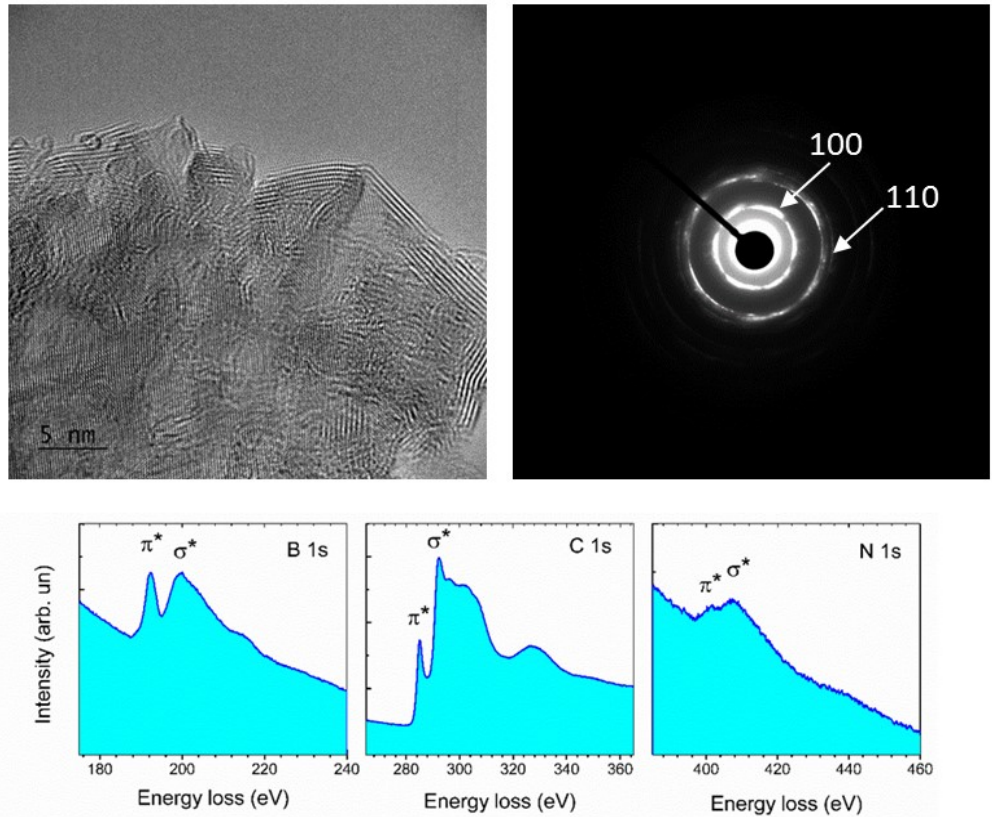

**Figure 4.** Top panel: TEM image of a bare B-C-N sample (left) and its corresponding SAED pattern (right). Diffraction planes of the hexagonal B-C-N structure corresponding to the observed diffraction rings are indicated. Bottom panel: EELS data showing the 1s core level signals of B, C and N.

The EELS spectrum of the compound as obtained by TEM in the 160 to 480 eV energy-loss region is reported in Figure 4. We clearly identify the three main energy-loss features associated to excitation from the B, C and N 1s core-levels to the first unoccupied orbitals. All EELS features present multi-peak structures. In particular, we observe the first (second) important K-edge absorption features at 190.8 eV (198.0 eV), 284.0 eV (291.5 eV) and 400.5 eV (406.5 eV) for B, C and N, respectively. These main and secondary absorption peaks can be associated to the typical $\pi^*$ and $\sigma^*$ resonances of the three elements [28-31]. We note the sharp and intense $\pi^*$ resonances, that demonstrate the dominance of $sp^2$-bonds throughout the sample, which would be absent in case of $sp^3$ hybridization [28,32,33]. Low loss spectrum (shown in the Supplementary Data, Fig.S5) is also similar to those of Gr and h-BN [34]. Like this, the formation of the $\pi^*$ band throughout the sample is unambiguously demonstrated, the planar hexagonal type bonding scheme is thus proved.

The XPS investigation of the B, C and N 1s core-levels at a B-C-N sample transferred onto Si substrate and annealed at 500 °C under ultra-high vacuum is presented in Figure 5. The core-level data present asymmetric shapes with tails towards high and low binding energies, suggesting the presence of a major component and side peaks for each core-level. We notice that the B, C and N core-levels in mixed ternary π-bonding compounds can produce asymmetric peaks with large tails rather than well-resolved peaks [10,12,49]. In order to evaluate the position and relative intensity of these contributions we performed a fitting analysis, whose fitting values are reported in the Supplementary Data (Table S3) and whose results with the deconvolution in different components contributing to the signals are also reported in Figure 5. The fitting has been carried out with pseudo-Voigt line shaped components (Gaussian and Lorentzian contributions, taking into account the experimental resolution and the intrinsic line shape of each core-level, respectively), after the subtraction of a Shirley-like background.

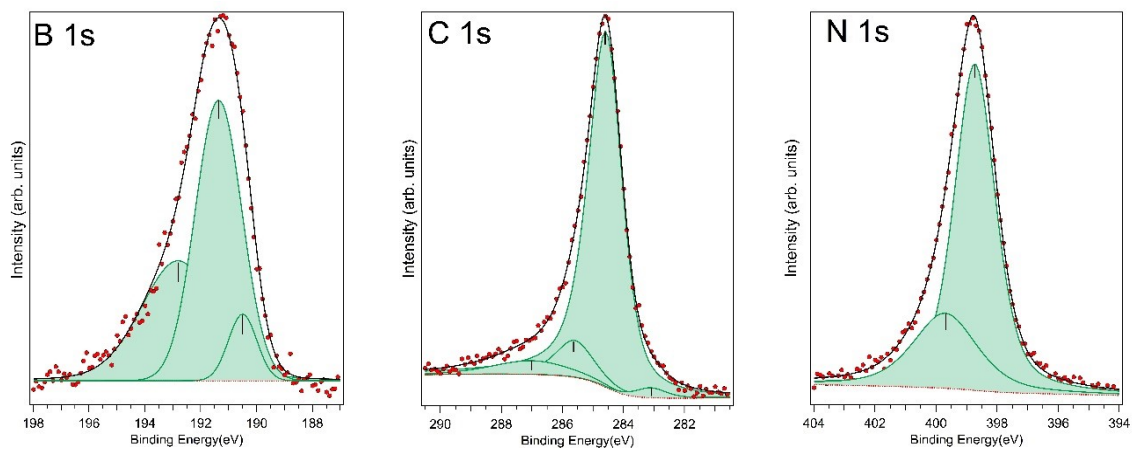

**Figure 5.** XPS peaks corresponding to B 1s, C 1s and N 1s core-levels of a B-C-N layer transferred onto Si substrate (experimental data, red dots), including the overall fitting curve (black lines) and fitting components (colored lines).

The C 1s core-level presents the dominant component associated to the planar sp$^2$-bonding configuration at 284.6 eV BE, which is in excellent agreement with previous observations on sp$^2$-bonded carbon, like planar B-C-N compounds [4,21,35-40] and Gr [41-43]. The C 1s line shape is an excellent fingerprint of the carbon atoms for a variety of systems on surfaces, giving direct information on their chemical environment [20,44-46]. A fitting analysis carried out using only this main peak with the expected parameters (slight asymmetry in the Lorentzian lineshape) is not sufficient to take into account the whole experimental C1s signal, though. Thus, we have to introduce other structures aside the main C 1s component, a further peak at higher energy (285.6 eV) and a small one at lower (283.1 eV) BE, that can be attributed to the interaction of carbon atoms with the more electronegative N atoms [4,36,47-49], and to carbon interaction with the less electronegative B [4,10,48,49], respectively. A less intense and broader peak at 287 eV BE is associated to residual oxygen and defects, reflecting the typical impurity broad multicomponent band present in graphite and carbon-based systems [50-54].

The N 1s signal presents the main structure at about 398.7 eV, associated to the N-B component typical of h-BN and B-C-N compounds [4,21,36,47,49,55,56]. Also for the N 1s core level, this main component is not sufficient to account for the whole spectral density. In particular, a broader peak at 399.7 eV must be introduced and associated to the N-C interaction, whose wider shape can be explained by the presence of N coordinated in different sites of the $sp^2$ matrix, where N can be substitutional with pyridinic, pyrrolic and graphitic sites [38,42,57,58]. The dominant peak of the B 1s core level at 191.4 eV can be associated to the main B-N interaction, but also two further components must be introduced to account for the low and high-BE spectral density: a small structure at 190.5 eV and a more intense and broad peak at 192.8 eV, the first one due to the presence of some B-C bonds [38,56,57], the second one due to a multicomponent B coordination to oxygen [59,60]. The most intense peak reflects the important coordination of B with nitrogen and a only minor component (of the order of 10%) with carbon. The peak due to B interacting with oxygen is fitted with a wide feature reflecting the multicomponent different stoichiometric environment because of the high reactivity of boron with oxygen with possible segregation of B-O groups at the sample surface [59,60].

The B, C and N 1s core-level data are consistent with the presence of dominant h-BN and Gr domains in the sample. However, there is an important presence of B and N interactions with C atoms that can be justified by the presence of B and N at the graphene borders, and of C interacting at the BN-edges, respectively, in the planar compound. These interactions do not exclude, though, the formation of high N-doping (and possibly some B doping too) of Gr-like domains and C-incorporation into h-BN, also consistent with the observed binding energies of the B, C and N core-levels. The core-level peak binding energy, full-width at half-maximum (FWHM) and relative intensity for each component of B 1s, C 1s and N 1s XPS peaks, estimated as the percentage of the total area normalized to the respective excitation cross section values [61], are included in the Supplementary Data (Table S3). The average B:C:N contents obtained by integrating the 1s core-level peaks taking into account the excitation cross sections [61], and excluding the oxygen components so to determine the only B-C N compound stoichiometry, are in a ratio 1:3:1.5. There is some excess of N atoms with respect to B that comes from the higher N-doping in the Gr-like domains and preferential C-N bonds at Gr/h-BN boundaries [7,10]. Therefore, the composition of the obtained B-C-N layers can be expressed as $(BN)_{0.25}$-$(C_{0.9}N_{0.1})_{0.75}$. Our results essentially agree with previous ones in the fact that segregation of C-rich (Gr-like) and h-BN domains takes place in the obtained layers [7,8,10,11]. However, our results show a significant C-N bonding coming from C-N interactions at the Gr/h-BN interfaces and to N doping in Gr domains. Previous work using DMAB as a single-source precursor [10] obtained samples with about 3% of B and N impurities in C-rich domains, whereas h-BN phase with a doping of 2-5% of C is got when using TMAB precursor [11]. Here, we obtained B-C-N samples with overall B:C:N contents in atomic proportions close to 1:3:1.5, showing about 10% of C atoms bound to N.

Raman spectroscopy is a widely used technique to characterize chemical bonding in 2D materials. Raman spectra of both Gr and h-BN present well defined characteristic features. Figure 6a plots the Raman spectrum of a layer transferred onto Si substrate, obtained under a laser excitation wavelength of 532 nm. The spectrum is quite similar to those obtained in previous works with Gr/hBN hybrid layers [7,8,40]. It presents a band at 1361 cm$^{-1}$, which can be ascribed to the contribution of D band related to defects in Gr [62] as well as the B-N stretching band [63]. The G Raman band typical of C-C bonds in Gr [62] is observed at 1598 cm$^{-1}$. Peak positions of these two bands exhibit an opposite shift by varying the laser excitation energy, as shown in Figure 6b. The observed D band dispersion is equal to 42±2 cm$^{-1}$/eV, in good agreement with the values reported for the D band in Gr and Carbon materials [62,64,65]. This observation indicates that the band at 1361 cm$^{-1}$ is essentially due to the presence of defects (i.e., to the D band), which implies that the B-N stretching contribution to that peak is negligible. Indeed, B-N stretching in h-BN is much weaker than the G band in Gr under the same excitation conditions [63], whereas the relative intensities of the present D and G bands are similar. As concerns the G band, it is also dispersive, showing a shift of -9±1 cm$^{-1}$/eV. This negative dispersion, which has been also reported in multilayered Gr and graphite samples [64,65], seems to be related to the presence of stacking faults between the layers [66]. On the other hand, the D and G band intensity ratios show a dispersion as a function of the excitation wavelength which is consistent with the assignment of the 1361 cm$^{-1}$ band to the D band [67], as shown in Figure 6c. The obtained defect concentration from this dispersion curve is equal to 1.8•10$^{11}$ cm$^{-2}$, which corresponds to a typical separation distance between point defects of about 13 nm [67]. The presence of high defect concentrations in the sample is confirmed by the shoulder appearing at 1631 cm$^{-1}$, which can be ascribed to D' band [7,8,40,68], which is also indicative of defects, and is usually observed in N-doped Gr [69]. This observation is consistent with the high N-doping of C-rich domains observed in XPS measurements as well as in the FTIR analyses of the samples (shown in the Supplementary Data, Fig.S9). The 2D and D+G overtones are also seen at 2714 and 2942 cm$^{-1}$. The ratio of the 2D and G relative intensities confirms that the samples are made of few layers instead of a monolayer [69]. All these bands are a clear signature of a hexagonal C-rich structure having a high concentration of defects. This picture is further supported by electrical conductance measurements (shown in the Supplementary Data, Fig.S10), giving a high electrical conductivity of 215 S•cm$^{-1}$ at room temperature and a low activation energy of 0.6±0.1 meV, which is presumably due to scattering of electrons with defects in the C-rich regions. It seems that these C-rich regions percolate through the sample, acting as pathways for electronic conduction through the layers.

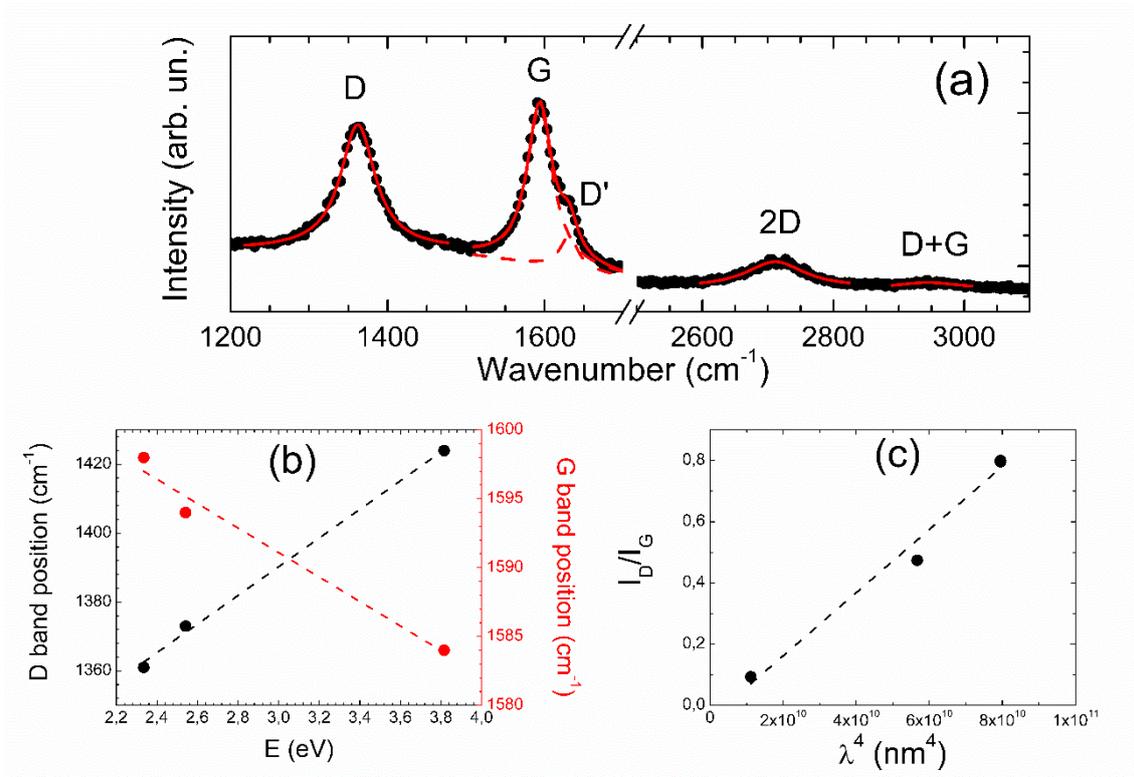

**Figure 6. (a)** Raman spectrum obtained under visible excitation (532 nm) of a B-C-N layer deposited onto Si substrate. The spectrum has been deconvoluted into several Lorentz peaks (indicated as D, G, D', 2D and D+G in the figure). **(b)** Dispersion of the D and G band positions as a function of the laser excitation wavelength. **(c)** Ratio of the intensities of the D and G bands as a function of the laser excitation wavelength to the fourth power.

**Optical bandgap determination**

The optical properties of the obtained layers have been investigated by UV-vis optical absorption spectroscopy. To this aim B-C-N layers have been transferred onto quartz substrates. The typical spectra of the bare quartz substrate and the B-C-N layer are plotted in Figure 7. It can be seen that quartz substrate has a flat signal in the investigated wavelength range. On the other hand, the B-C-N layer presents two clear absorption bands with maxima at 270 and 200 nm, respectively. A very common way of analyzing optical absorption bands and determining optical bandgaps is by means of Tauc's theory. According to Tauc's seminal work [70], the bandgap of an indirect allowed transition can be obtained from the following relationship:

$$\omega^2\varepsilon=(\hbar\omega-E_g)^2 \qquad (2)$$

where $\omega$ is the angular frequency of the light, $E_g$ is the bandgap energy, $\hbar$ is the Planck's constant and $\varepsilon$ is the imaginary part of the dielectric constant. In some previous seminal works on layers of similar compositions [7,8,71,72] the authors stated that $\varepsilon$ is the optical absorbance, a concept which is somewhat diffuse. It rather seems that in these works it was considered that $\varepsilon$ is equivalent to the optical absorption coefficient ($\alpha$) of the material. Actually, if we plot our experimental data in that way, we obtain very similar results to those previously reported (see the Supplementary Data, Fig.S11). This analysis gives optical bandgaps in our samples of about 1.7 eV and 5.2 eV, which are in good accordance with previously reported values in Gr/hBN hybrid layers [7,8,72]. On the basis of these results some authors claimed that B and N doping can open a large bandgap of 1.7 eV in Gr-like domains and that C-doping reduced the bandgap of h-BN domains. These are significant results with important consequences. However, the former analysis relies on an erroneous interpretation of Tauc's theory. Actually, the proper analysis of the absorption spectra must be done by considering the relationship between the imaginary part of the dielectric constant appearing in Tauc's equation ($\varepsilon$) and the optical absorption coefficient ($\alpha$) or the optical density (O.D.), which are the most usual parameters determined experimentally. This analysis gives a very different picture to that exposed above (see the details in the Supplementary Data, Fig.S12). The plots of the experimental UV-vis data obtained with B-C-N films according to Tauc's method give optical bandgaps of 3.4±0.1 and 5.8±0.2 eV. The bandgap of 5.8 eV agree with the reported optical bandgap of h-BN [11,73-75]. Therefore, the optical absorption band with a peak at 200 nm is ascribed to h-BN domains in the B-C-N layers. On the other hand, the absorption band showing a maximum at 270 nm is usually observed in Gr-like samples. It must be noticed that this transition cannot be properly modelled by Tauc's theory and, therefore the obtained bandgap of 3.4 eV can be only taken as an indicative or reference value to compare absorption spectra of Gr-like samples analyzed by the same method. Actually, this optical absorption band can be properly described by the Fano model [76]. The fit of experimental optical absorption data with this model is included in Figure 7. The obtained fitting parameters are $E_0$ = 5.20 eV (the band to band transition energy at the M point singularity predicted from ab initio GW calculations), $E_{res}$ = 4.95 eV (the resonance energy of the perturbed exciton), $\Gamma$ = 1.56 eV (width relative to the resonance energy $E_{res}$ of the perturbed exciton) and q = −1 ($q^2$ defines the ratio of the strength of the excitonic transition to the unperturbed band transitions, while the sign of q determines the asymmetry of the line shape). As compared to the values reported for monolayer Gr samples ($E_0$ = 5.20 eV, $E_{res}$ = 5.02 eV, $\Gamma$ = 0.78 eV, q = -1) [76], we obtain a slightly lower resonance energy and a higher width. Such effects can be tentatively ascribed to layer thickness and doping effects [76,77] as well as defect concentration. In resume, the results of the optical absorption measurements clearly confirm that the sample is composed of two separate phases, namely hybridized C-rich and BN-rich domains, in good accordance with XPS results, FTIR and Raman signatures.

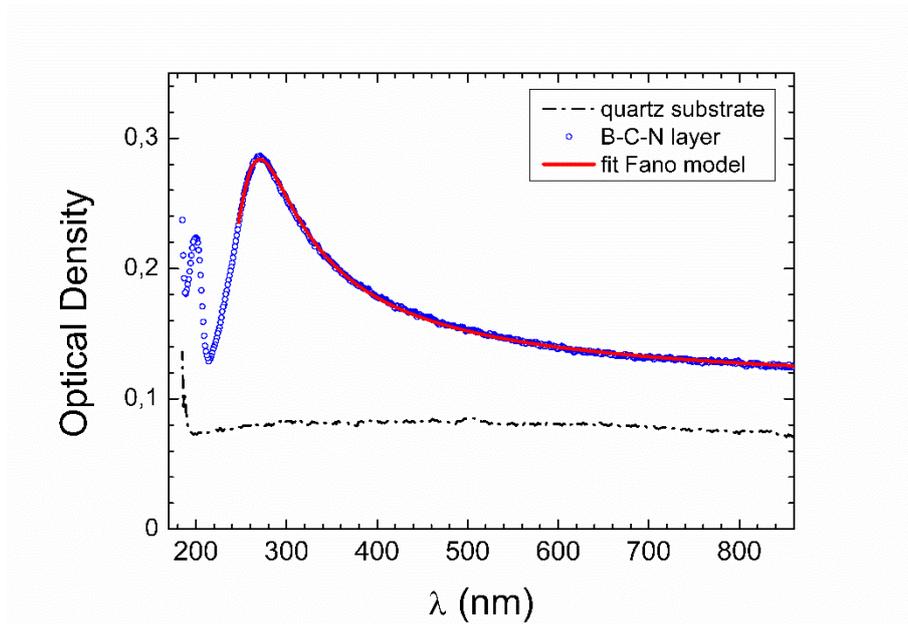

**Figure 7.** Typical optical absorption spectra of the obtained B-C-N layers (blue circles) and of the quartz substrate (black crosses). The red line is the fit of experimental absorption band at 270 nm corresponding to C-rich domains by using the Fano model.

**Conclusions**

The use of Methylamine Borane (MeAB) as a single-source precursor for the CVD growth of B-C-N layers is reported. The gaseous products released during the thermolysis of MeAB have been determined by mass spectrometry analyses. Thermolytic decomposition of MeAB under the CVD conditions used here leads to the evolution of gaseous MeAB and cyclic trimethylborazine molecules, which are the actual precursors for B-C-N growth on Cu substrates at 1000 ºC. This synthesis path gives rise to the formation of Gr-like and h-BN nanodomains in the obtained B-C-N layers. The overall B:C:N contents in the samples are in atomic proportions close to 1:3:1.5, showing about 10% of C atoms bound to N. Furthermore, it is also highlighted how the claiming of an optical band gap opening in Gr-rich domains and a significant reduction of h-BN bandgap reported in previous literature is based on an incorrect analysis of the optical absorption spectra, and how a proper analysis offers a quite different description of the optical properties of Gr/h-BN hybrids. A detailed analysis of the optical bandgaps in these hybrid layers has been done on the basis of Tauc's and Fano's models, which are the proper descriptions to depict the optical transitions of h-BN and Gr-like domains, respectively.

**Acknowledgements**

This work has been funded under "EXPLORA" Program of Spanish MINECO (grant number FIS2014-61634-EXP) and the MINECO contract MAT2014-53119-C2-1-R. Part of this work (the synthesis and characterization of MeAB) was also supported by COST Action MP1103 "Nanostructured materials for solid-state hydrogen storage". Technical assistance from Mr. Fernando Moreno and Segainvex Facilities at UAM is also gratefully acknowledged.

**Supplementary Data**

Results of XRPD, NMR and FTIR characterization of MeAB precursor, crystallographic data of MeAB, mass spectra and isothermal desorption profiles of MeAB thermolysis products, results of EELS, XPS, Raman, FTIR, electrical conductivity and optical bandgap determination of the obtained B-C-N layers.

**References**

[1] da Rocha Martins, J.; Chacham, H. "Disorder and Segregation in B-C-N Graphene-Type Layers and Nanotubes: Tuning the Band Gap", ACS Nano **2011**, 5, 385–393

[2] Wang, S.; Zhang, L.; Xia, Z.; Roy, A.; Chang, D.W.; Baek, J.-B.; Dai, L. "BCN Graphene as Efficient Metal-Free Electrocatalyst for the Oxygen Reduction Reaction", Angew. Chem. Int. Ed. **2012**, 51, 4209 –4212

[3] Chhetri, M.; Maitra, S.; Chakraborty, H.; Waghmareb, U.V.; Rao, C.N.R. "Superior performance of borocarbonitrides, BxCyNz, as stable, low-cost metal-free electrocatalysts for the hydrogen evolution reaction", Energy Environ. Sci., **2016**, 9, 95

[4] Huang, C.; Chen, C.; Zhang, M.; Lin, L.; Ye, X.; Lin, S.; Antonietti, M.; Wang, X. "Carbon-doped BN nanosheets for metal-free photoredox catalysis", Nature Comm. **2015**, 6:7698

[5] Li,M.; Wang, Y.; Tang,P.; Xie, N.; Zhao, Y.; Liu, X.; Hu, G.; Xie,J.; Zhao, Y.; Tang, J.; Zhang, T.; Ma, D. "Graphene with Atomic-Level In-Plane Decoration of h-BN Domains for Efficient Photocatalysis" Chem. Mater. **2017**, 29, 2769-2776

[6] Guilhon, I.; Marques, M.; Teles, L. K. "Optical absorbance and band-gap engineering of (BN)1−x(C2)x two-dimensional alloys: Phase separation and composition fluctuation effects", Phys. Rev. B **2017** 95, 035407

[7] Ci, L.; Song, L.; Jin, C.; Jariwala, D.; Wu, D.; Li, Y.; Srivastava, A.; Wang, Z. F.; Storr, K.; Balicas, L.; Liu, F.; Ajayan, P. M. Atomic Layers of Hybridized Boron Nitride and Graphene Domains. Nature Mater. **2010**, 9, 430−435.

[8] Chang, C.-K.; Kataria, S.; Kuo, C.-C.; Ganguly, A.; Wang, B.-Y.; Hwang, J.-Y.; Huang, K.-J.; Yang, W.-H.; Wang, S.-B.; Chuang, C.-H.; Chen, M.; Huang, C.-I.; Pong, W.-F.; Song, K.-J.; Chang, S.-J.; Guo, J.-H.; Tai, Y.; Tsujimoto, M.; Isoda, S.; Chen, C.-W.; Chen, L.-C.; Chen, K.-H. Band Gap Engineering of Chemical Vapor Deposited Graphene by in Situ BN Doping. ACS Nano **2013**, 7, 1333−1341

[9] Levendorf, M.P.; Kim, C.-J.; Brown, L.; Huang, P.Y.; Havener, R.W.; Muller, D.A.; Park J. "Graphene and boron nitride lateral heterostructures for atomically thin circuitry", Nature **2012**, 488, 627-632

[10] Nappini, S.; Píš, I.; Mentes, T.O.; Sala, A.; Cattelan, M.; Agnoli, S.; Bondino, F.; Magnano, E. "Formation of a Quasi-Free-Standing Single Layer of Graphene and Hexagonal Boron Nitride on Pt(111) by a Single Molecular Precursor", Adv. Funct. Mater. **2015**, 26, 1120-1126

[11] Tay, R.Y.; Li, H. Tsang, S.H.; Zhu, M.; Loeblein, M.; Jin, L.; Leong, F.N.; Teo, E.H.T. "Trimethylamine Borane: A New Single-Source Precursor for Monolayer h‑BN Single Crystals and h‑BCN Thin Films", Chem. Mater. **2016**, 28, 2180−2190

[12] Beniwal, S.; Hooper, J.; Miller, D.P.; Costa, P.S.; Chen, G.; Liu, S.-Y.; Dowben, P.A.; Sykes, E.C.H.; Zurek, E.; Enders, A. "Graphene-like Boron−Carbon−Nitrogen Monolayers", ACS Nano **2017**, 11, 2486−2493

[13] Bowden, M.E.; Brown, I.W.M.; Gainsford, G.J.; Wong, H. 'Structure and thermal decomposition of methylamine borane', Inorg. Chim. Acta **2008**, 361, 2147-2153

[14] Yang, L.; Su, J.; Meng, X.; Luo, W.; Cheng, G. 'In situ synthesis of graphene supported Ag@CoNi core-shell nanoparticles as highly efficient catalysts for hydrogen generation from hydrolysis of ammonia borane and methylamine borane', J. Mater. Chem. A, **2013**, 1, 10016-10023

[15] Beachley, 0.T. "Intermediates in the Formation of N-Methylaminoborane Trimer and N,N-Dimethylaminoborane Dimer", Inor. Chem. **1967**, 6, 870-874

[16] Boultif, A.; Louer, D. "Indexing of powder diffraction patterns for low-symmetry lattices by the successive dichotomy method", J. Appl. Cryst. **1991**, 24, 987-993

[17] Favre-Nicolin, V.; Černý, R. "FOX, `free objects for crystallography': a modular approach to ab initio structure determination from powder diffraction", J. Appl. Crystallogr. **2002**, 35, 734-743

[18] Petricek, V.; Dusek, M.; Palatinus, L. "Crystallographic Computing System JANA2006: General features" Zeitschrift für Kristallographie - Crystalline Materials, **2014**, 229, 345-352

[19] Horcas, I.; Fernández, R.; Gómez-Rodríguez, J.M.; Colchero, J.; Gómez-Herrero, J.; Baro, A.M. "WSXM: A software for scanning probe microscopy and a tool for nanotechnology", Review of Scientific Instruments, **2007**, 78, 013705

[20] Massimi, L.; Angelucci, M.; Gargiani, P.; Betti, M.G.; Montoro, S. "Metal-phthalocyanine ordered layers on Au(110): Metal-dependent adsorption energy", The Journal of Chemical Physics, **2014**, 140, 244704

[21] Massimi, L.; Betti, M.G.; Caramazza, S.; Postorino, P.; Mariani, C.; Latini, A.; Leardini, F. "In-Vacuum Thermolysis of Ethane 1,2-Diamineborane for the Synthesis of Ternary Borocarbonitrides", Nanotechnology, **2016**, 27, 435601

[22] Abrahams, S. C.; Kalnajs, J. "The Lattice Constants of the Alkali Borohydrides and the Low-Temperature Phase of Sodium Borohydride", J. Chem. Phys., **1954**, 22, 434-436

[23] Framery, E.; Vaultier, M. "Efficient Synthesis and NMR Data of N- or B Substituted Borazines", Heteroatom Chemistry, **2000**, 11, 218-225

[24] Yamamoto, Y.; Miyamoto, K.; Umeda, J.; Nakatani, Y.; Yamamoto, T.; Miyaura, N. "Synthesis of B-trisubstituted borazines via the rhodium-catalyzed hydroboration of alkenes with N,N′,N″-trimethyl or N,N′,N″-triethylborazine" J. Organometallic Chemistry, **2006**, 691, 4909–4917

[25] Kawaguchi, M.; Kawashima, T.; Nakajima, T. "Syntheses and Structures of New Graphite-like Materials of Composition BCN(H) and BC3N(H)", Chem. Mater. **1996**, 8(1996) pp. 1197-1201

[26] Williams, D.B.; Carter, C.B. "Transmission Electron Microscopy: A Textbook for Materials Science" **2009**, Spinger, 2$^{nd}$ Edition

[27] Karmakar S.; Nawale, A.B.; Kanhe, N.S.; Sathe, V.G. Mathe, V.L.; Bhoraskar, S.,V. "Tailored conversion of synthetic graphite into rotationally misoriented few-layer graphene by cold thermal shock driven controlled failure" Carbon **2014**, 67, 534-545

[28] Schmid, H.K. "Phase Identification in Carbon and BN Systems by EELS", Microsc. Microanal. Microstruct. **1995**, 6, 99-111

[29] Arenal, R.; Kociak, M.; Zaluzec, N.J. "High-angular-resolution electron energy loss spectroscopy of hexagonal boron nitride" Appl. Phys. Lett. **2007**, 90, 204105

[30] Lu, J.; Gao, S.-P.; Yuan, J. "ELNES for boron, carbon, and nitrogen K-edges with different chemical environments in layered materials studied by density functional theory" Ultramicroscopy **2012**, 112, 61-68

[31] Wibbelt, M.; Kohl, H.; Kohler-Redlich, Ph. "Multiple-scattering calculations of electron-energy-loss near-edge structures of existing and predicted phases in the ternary system B-C-N" Phys. Rev. B **1999**, 59, 11739

[32] McCulloch, D.G.; Lau, D.W.M.; Nicholls, R.J.; Perkins, J.M. "The near edge structure of cubic boron nitride" Micron **2012**, 43, 43-48

[33] Meng, Y.; Mao, H.-K.; Eng, P.J.; Trainor, T.P.; Newville, M.; Hu, M.Y.; Kao, C.; Shu, J.; Hausermann, D.; Hemley, R.J. "The formation of sp3 bonding in compressed BN" Nature Materials **2004**, 3, 111-114

[34] Egerton, R.F. "Electron Energy-Loss Spectroscopy in the Electron Microscope", Springer **2011**, Third Edition ISBN 978-1-4419-9582-7

[35] Sulyaeva, V. S.; Kosinova, M. L.; Rumyantsev, Y.M; Golubenko, A.N.; Fainer, N.I.; Alferova, N.I.; Ayupov, B.M.; Gevko, P.N.; Kesler, V.G.; Kolesov, B.A.; Maksimovskii, E.A.; Myakishev, K.G.; Yushina, I.V.; Kuznetsov, F.A. "Properties of BC$_x$N$_y$ Films Grown by Plasma-Enhanced Chemical Vapor Deposition from N-Trimethylborazine–Nitrogen Mixtures", Inorganic Materials, **2010**, 46, 5, 487-494

[36] Sulyaeva, V.S.; Kosinova, M.L.; Rumyantsev, Y.M.; Kesler, V.G.; Kuznetsov, F.A. "PECVD synthesis and optical properties of BCXNY films obtained from N-triethylborazine as a single-source precursor"; Surface & Coatings Technology **2013**, 230, 145–151

[37] Sulyaeva, V.S.; Kosinova, M.L.; Rumyantsev, Y.M.; Kuznetsov, F.A.; Kesler, V.G.; Kirienko, V.V. "Optical and electrical characteristics of plasma enhanced chemical vapor deposition boron carbonitride thin films derived from N-trimethylborazine precursor" Thin Solid Films **2014**, 558, 112-117

[38] Kang, Y.; Chu, Z.; Zhang, D.; Li, G.; Jiang, Z.; Cheng, H.; Li, X. "Incorporate boron and nitrogen into graphene to make BCN hybrid nanosheets with enhanced microwave absorbing properties" Carbon **2013**, 61, 200-208

[39] Chiang, W.-H.; Hsieh, C.-Y., Lo, S.-C.; Chang, Y.-C.; Kawai, T.; Nonoguchi, Y. "C/BCN core/shell nanotube films with improved thermoelectric properties" Carbon **2016**, 109, 49-56

[40] Wang, H.; Zhao, C.; Liu, L.; Xu, Z.; Wei, J.; Wang, W.; Bai, X.; Wang, E. "Towards the controlled CVD growth of graphitic B–C–N atomic layer films: The key role of B–C delivery molecular precursor" Nano Res. **2016**, 9, 1221-1235

[41] Usachov D, Vilkov O, Gruneis A, Haberer D, Fedorov A, Adamchuk V, Preobrajenski A, Dudin P, Barinov A, Oehzelt M.; Laubschat, C.; Vyalikh, D.V.

Nitrogen-Doped Graphene: Efficient Growth, Structure, and Electronic Properties" Nano letters **2011**, 11 5401–5407

[42] Scardamaglia, M.; Lisi, S.; Lizzit, S.; Baraldi, A.; Larciprete, R.; Mariani, C.; Betti, M.G. "Graphene-Induced Substrate Decoupling and Ideal Doping of a Self-Assembled Iron-phthalocyanine Single Layer" J. Phys. Chem. C **2013**, 117, 3019- 3027

[43] Gupta, B.; Bernardo, I.D.; Mondelli, P.; Pia, A.D.; Betti, M.G.; Iacopi, F.; Mariani, C.; Motta, N. "Effect of substrate polishing on the growth of graphene on 3C–SiC(111)/Si(111) by high temperature annealing" Nanotechnology **2016**, 27, 185601

[44] Massimi, L.; Ourdjini, O.; Laffrentz, L.; Koch, M.; Grill, L.; Cavaliere, E.; Gavioli, L.; Cardoso, C.; Prezzi, D.; Molinari, E.; Ferretti. A.; Mariani, C.; Betti, M.G. "Surface-Assisted Reactions toward Formation of Graphene Nanoribbons on Au(110) Surface", J. Phys. Chem. C **2015**, 119, 2427–2437

[45] Baldacchini, C.; Allegretti, F.; Gunnella, R.; Betti, M.G. "Molecule–metal interaction of pentacene on copper vicinal surfaces" Surf. Sci. **2007**, 601, 2603-2606

[46] Avvisati, G.; Lisi, S.; Gargiani, P.; Pia, A.D.; De Luca, O.; Pacilé, D.; Cardoso, C.; Varsano, D.; Prezzi, D.; Ferrett, A.; Betti, M.G. "FePc Adsorption on the Moiré Superstructure of Graphene Intercalated with a Cobalt Layer" J. Phys. Chem. C, **2017**, 121, 1639- 1647

[47] Huang, F.L.; Cao, C.B.; Xiang, X.; Lv, R.T.; Zhu, H.S. "Synthesis of hexagonal boron carbonitride phase by solvothermal method" Diamond & Related Materials **2004**, 13, 1757-1760

[48] Wang, L.; Wu, J.; Wang, L.; Guo, C.; Xu, Y. "High-yield synthesis of uniform B, N-rich BN-Cx nanoplates in mild temperatures" J Nanopart. Res. **2014**, 16, 2511

[49] Sreedhara, M. B.; Gopalakrishnan, K.; Bharath, B.; Kumar, R.; Kulkarni, G.U.; Rao, C.N.R. "Properties of Nanosheets of 2D Borocarbonitrides Related to Energy Devices, Transistors and Other Areas" Chem. Phys. Lett. **2016**, 657, 124−130

[50] Dose, V.; Reusing, G.; Scheidt, H. "Unoccupied electronic states in graphite" Physical Review B **1982**, 26, 984

[51] Yue, Z.; Jiang, W.; Wang, L.; Gardner, S.; Pittman, C. "Surface characterization of electrochemically oxidized carbon fibers", Carbon **1999**, 37, 1785-1796

[52] Estrade-Szwarckopf, H. "XPS photoemission in carbonaceous materials: A "defect" peak beside the graphitic asymmetric peak" Carbon 2004, 42 1713-1721

[53] Yang, D.Q.; Sacher, E. "Carbon 1s X-ray Photoemission Line Shape Analysis of Highly Oriented Pyrolytic Graphite: The Influence of Structural Damage on Peak Asymmetry" Langmuir **2006**, 22 860-862

[54] Stankovich, S.; Dikin, D.A.; Piner, R.D.; Kohlhaas, K.A.; Kleinhammes, A.; Jia, Y.; Wu, Y.; Nguyen, S.T.; Ruoff, R.S. "Synthesis of graphene-based nanosheets via chemical reduction of exfoliated graphite oxide", Carbon **2007**, 45, 1558-1565

[55] Henck, H, Pierucci, D, Ben Aziza, Z, Silly, MG, Gil, B, Sirotti, F, Cassabois, G, Ouerghi, A. "Stacking fault and defects in single domain multilayered hexagonal boron nitride" Appl. Phys. Lett. **2017**, 110, 023101 (2017)

[56] Kumar, N.; Moses, K.; Pramoda, K.; Shirodkar, S.N.; Mishra, A.K.; Waghmare, U.V.; Sundaresan, A.; Rao, C.N.R. "Borocarbonitrides, $B_xC_yN_z$" J. Mater. Chem. A **2013**, 1, 5806-5821

[57] Moses, K.; Shirodkar, S.N.; Waghmare, U.V.; Rao, C.N.R. "Composition-dependent photoluminescence and electronic structure of 2-dimensional borocarbonitrides, $BC_xN$ (x = 1, 5)" Materials Research Express **2014**, 1, 025603

[58]    Scardamaglia, M.; Struzzi, C.; Aparicio Rebollo, F.J.; De Marco, P.; Mudimela, P.R.; Colomer, J.-F.; Amati, M.; Gregoratti, L.; Petaccia, L.; Snyders, R. "Tuning electronic properties of carbon nanotubes by nitrogen grafting: Chemistry and chemical stability" Carbon **2015**, 83, 118 – 127

[59] Ong, C.W.; Huang, H.; Zheng, B.; Kwok, R.W.M.; Hui, V; Lau, W.M. "X-ray photoemission spectroscopy of nonmetallic materials: Electronic structures of boron and $B_xO_y$" J. Appl. Phys. **2004**, 95, 3527

[60]    Paul, R.; Voevodin, A.A.; Zemlyanov, D.; Roy, A.K.; Fisher, T.S. "Microwave-Assisted Surface Synthesis of a Boron–Carbon–Nitrogen Foam and its Desorption Enthalpy" Adv. Funct. Mater. **2012**, 22, 3682-3690

[61] Yeh, J.J.; Lindau, I. "Atomic subshell photoionization cross sections and asymmetry parameters: $1 \leqslant Z \leqslant 103$" Atomic Data and Nuclear Data Tables **1985**, 32, 1-155

[62] Ferrari, A.C. "Raman spectroscopy of graphene and graphite: Disorder, electron–phonon coupling, doping and nonadiabatic effects", Solid State Communications **2007**, 143, 47–57

[63] Gorbachev, R.V.; Riaz, I.; Nair, R.R.; Jalil, R.; Britnell, L.; Belle, B.D.; Hill, E.W.; Novoselov, K.S.; Watanabe, K.; Taniguchi, T.; Geim, A.K.; Blake, P. "Hunting for Monolayer Boron Nitride: Optical and Raman Signatures", Small **2011**, 7, 465–468

[64] Kaniyoor, A.; Ramaprabhu, S. "A Raman spectroscopic investigation of graphite oxide derived graphene", AIP Advances **2012**, 2, 032183

[65] Wang, Y.; Alsmeyer, D.C.; McCreery, R.L. "Raman Spectroscopy of Carbon Materials: Structural Basis of Observed Spectra", Chem. Mater. **1990**, 2, 557-563

[66] Ni, Z.; Liu, L.; Wang, Y.; Zheng, Z.; Li, L.-J.; Yu, T.; Shen, Z. "G-band Raman double resonance in twisted bilayer graphene: Evidence of band splitting and folding" Phys. Rev. B **2009**, 80, 125404

[67] Cancado, L. G.; Jorio, A.; Martins Ferreira, E.H.; Stavale, F.; Achete, C.A.; Capaz, R.B.; Moutinho, M.V.O.; Lombardo, A.; Kulmala, T.S.; Ferrari, A.C. "Quantifying Defects in Graphene via Raman Spectroscopy at Different Excitation Energies", Nano Lett. **2011**, 11, 3190–3196

[68] Pimenta, M.A.; Dresselhaus G, Dresselhaus MS, Cançado LG, Jorio A, Saito R. et al. "Studying disorder in graphite-based systems by Raman spectroscopy" Phys. Chem. Chem. Phys. **2007**, 9, 1276-1291

[69] Sun, Z.; Yan, Z.; Yao, J.; Beitler, E.; Zhu, Y.; Tour, J.M. "Growth of graphene from solid carbon sources, Nature **2010**, 468, 549

[70] Tauc, J.; Grigorovici, R.; Vancu, A. "Optical Properties and Electronic Structure of Amorphous Germanium" Phys. Stat. Sol. **1966**, 15, 627-637

[71] Song, L.; Ci, L.; Lu, H.; Sorokin, P.B.; Jin, C.; Ni, J.; Kvashnin, A.G.; Kvashnin, D.G.; Lou, J.; Yakobson, B.I.; Ajayan, P.M. "Large Scale Growth and Characterization of Atomic Hexagonal Boron Nitride Layers" Nano Lett. **2010**, 10, 3209-3215

[72] Kang, Y.; Chu, Z.; Zhang, D.; Li, G.; Jiang, Z.; Cheng, H.; Li, X. "Incorporate boron and nitrogen into graphene to make BCN hybrid nanosheets with enhanced microwave absorbing properties" Carbon **2013**, 61, 200-208

[73] Watanabe, K.; Taniguchi, T.; Kanda, H. "Direct-bandgap properties and evidence for ultraviolet lasing of hexagonal boron nitride single crystal", Nature Mat.**2004**, 3, 404-409

[74] Cassabois, G.; Valvin, P.; Gil, B. "Hexagonal boron nitride is an indirect bandgap semiconductor" Nature Photonics, **2016**, 10, 262-267

[75] Jalaly, M.; Gotor, F.J.; Semnan, M.; Sayagués, M.J. "A novel, simple and rapid route to the synthesis of boron cabonitride nanosheets: combustive gaseous unfolding" Sci. Rep. **2017**, 7, 3453

[76] Mak, K.F.; Shan, J.; Heinz, T.F.; "Seeing Many-Body Effects in Single- and Few-Layer Graphene: Observation of Two-Dimensional Saddle-Point Excitons", Phys. Rev. Lett. **2011**, 106, 046401

[77] Mak, K.F.; da Jornada, F.H.; He, K.; Deslippe, J.; Petrone, N.; Hone, J.; Shan, J.; Louie, S.G.; Heinz, T.F. "Tuning Many-Body Interactions in Graphene: The Effects of Doping on Excitons and Carrier Lifetimes", Phys. Rev. Lett. **2014**, 112, 207401